\documentclass[twocolumn,prd]{revtex4}% 
\pdfoutput=1

\usepackage{epsfig}
\usepackage{graphics}

\begin{document}
\bigskip
\hskip 4in\vbox{\baselineskip12pt \hbox{FERMILAB-PUB-12-044-A}  }
\bigskip\bigskip\bigskip

\title{A  Model of Macroscopic Geometrical  Uncertainty}

\author{Craig J. Hogan}
\affiliation{  University of Chicago and  Fermilab}

\begin{abstract}  

A   model  quantum system is proposed  to describe position states of a massive body in  flat space on large scales, excluding all standard quantum and gravitational degrees of freedom.  The model is based on  standard quantum spin commutators, with operators  interpreted as positions instead of  spin, and   a  Planck-scale length   $\ell_P$ in place of Planck's constant $\hbar$.  The algebra is used to derive a new  quantum geometrical uncertainty in direction, with variance given by $\langle \Delta \theta^2\rangle =   \ell_P/L$ at separation $L$, that dominates over standard quantum position uncertainty for bodies greater than the Planck mass.     The  system is discrete and holographic, and agrees with  gravitational entropy if the commutator coefficient takes the exact value $\ell_P= l_P/\sqrt{4\pi}$, where $l_P\equiv \sqrt{\hbar G/c^3}$ denotes the standard Planck length.  A physical interpretation is proposed that connects the operators with properties of classical position in the macroscopic limit:  Approximate locality and causality emerge in  macroscopic systems if position states  of multiple bodies are entangled  by proximity.  This interpretation predicts coherent directional fluctuations with variance $\langle \Delta \theta^2\rangle $ on timescale $\tau \approx L/c$ that lead to precisely predictable correlations in signals between adjacent  interferometers. It is argued that such a signal could provide compelling evidence of Planck scale quantum geometry, even in the absence of a complete dynamical or fundamental theory.
\end{abstract}
\maketitle
\section{Introduction}

In general relativity, space-time is a  dynamical classical system that exchanges energy and information with classical matter.  Of course, experiments prove\cite{ zeilinger1999,Ma2012,megidish} that matter is really a quantum system whose states are, in general, spatially delocalized\cite{EPR,cat,wigner,salecker}.
It is  not known how a determinate dynamical  geometry responds to the spatial indeterminacy of matter quantum states. 

Quantum field theory\cite{wilczek1999}  partially resolves this paradox by quantizing field modes in a classical geometry. The system of fields is a non localized entity of plane waves whose spatial structure is classical and determinate.  Interactions of fields are localized , but a measurement or interaction at a localized place causes a change of state of the field system which is nonlocal. This framework succeeds 
for particle interactions, but does not lead to a consistent quantum theory of dynamical geometry.

 One resolution of this long-standing tension could be  that  
 space-time geometry is also a quantum system, and  that classical features, such as locality, only emerge as  approximate macroscopic behaviors\cite{Banks:2011av}. This idea appears to be consistent with the observed classical dynamical behavior of  bodies, and indeed it has been shown that classical motion, Newtonian gravity, and general relativity can be derived from a statistical  theory\cite{Jacobson:1995ab,Verlinde:2010hp}.    
On the other hand the detailed character of the states of  quantum geometry, and   the  fidelity of the classical approximation to this statistical behavior,  are not known. 

This paper presents a macroscopic quantum model of  geometrical states--- an effective description of new geometrical degrees of freedom of position. 
 It starts with an approximation,  based on a classical massive body embedded in a quantum geometry,   complementary to quantum field theory, which is based on  quantized  matter fields  embedded in a classical background geometry. 
The model here cannot describe elementary particles--- indeed, it does not even have any dynamics--- but offers a candidate quantum description for  macroscopic geometrical states, and the departures they generate from classical behavior.

The model  here offers a new description of a very simple system: the position of a distant, massive body at rest in flat space-time. In standard physics this system is completely trivial--- quantum effects are negligible, and nothing at all happens classically. However, in the   framework proposed here,  where macroscopic directional information is limited, the new geometrical degrees of freedom display a new form of observable, coherent uncertainty and fluctuation in direction, whose properties can be related to classical gravity, and possibly even be observed.

\subsection{Motivation and Context}

The model described here is not a fundamental theory, but an effective or emergent description that applies only on large scales. We conjecture that the residual Planckian quantum effects of geometry on large scales can be described by certain directional degrees of freedom with specific properties.  In this spirit, it is  similar to  quantum descriptions  of collective excitations in condensed matter systems (e.g., ref. \cite{kohn99,laughlin99}).
  
 \subsubsection{Quantum Character of Dynamical Geometry}
 
Since dynamical geometry  exchanges energy and information with quantum matter, it must itself also be a quantum system.
Several  hints from theory suggest that  that we should consider the possibility of new, small but non negligible  of quantum behaviors of geometry, even on large scales.
  
\begin{figure}%[htbp]
 \epsfysize=2.5in 
\epsfbox{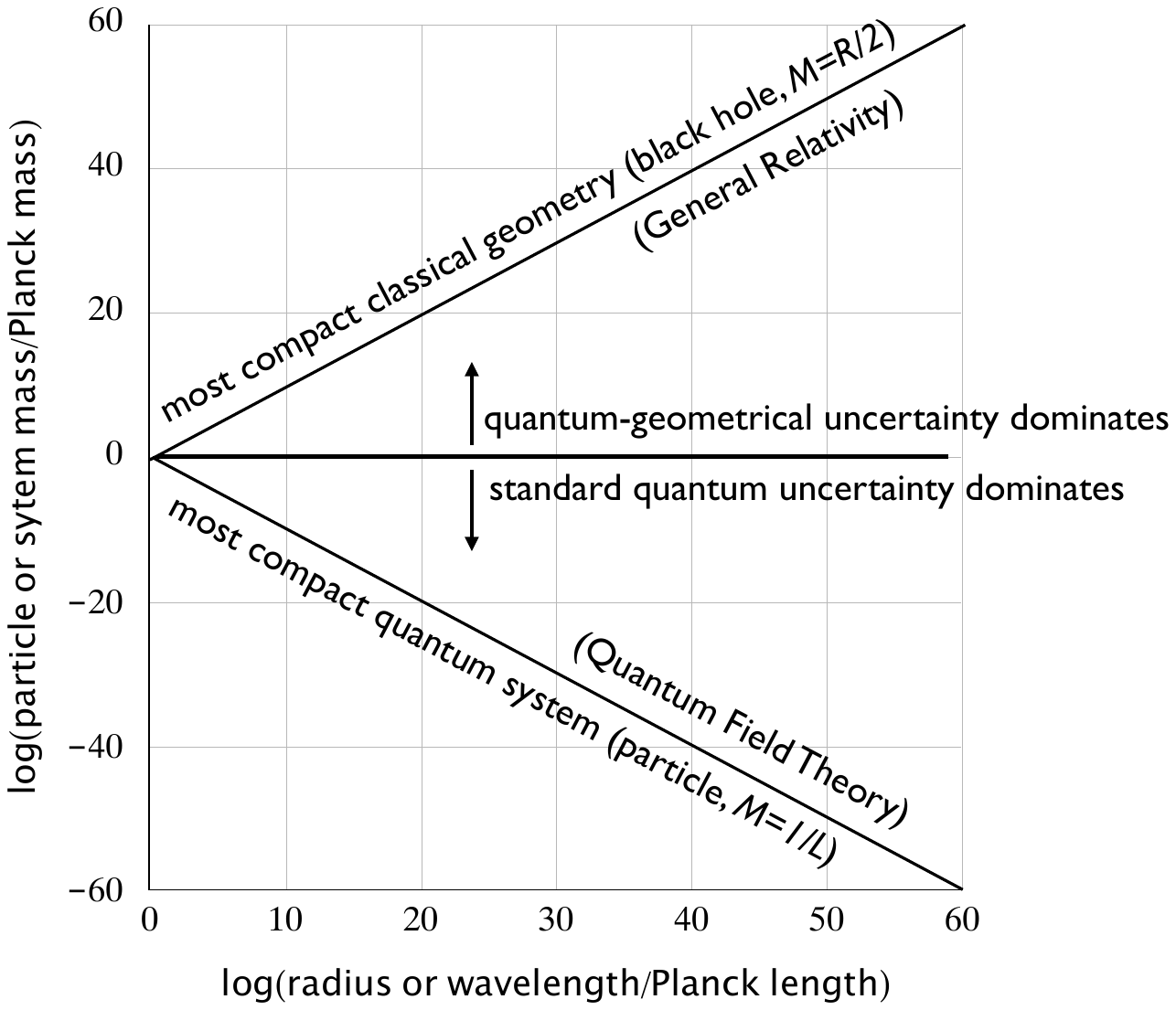} 
\caption{ \label{domains}
Domains of applicability of  approximations, in terms of the logarithm of the size and energy of a physical system in Planck units. The photoelectric relation in the lower half relates the wavelength and energy of a single  quantum. More compact systems  do not exist, since quanta  do not come in smaller packets. The Schwarzschild formula in the upper half relates the radius and mass of a black hole. Again, more compact systems do not exist, since a black hole is the most compact configuration of space-time for a given mass. All physical systems lie between these lines.  The the left of these two lines, in particular in a system smaller than  the Planck length where the two lines meet,  no system based on classical dynamical geometry responding to quantum matter can exist.
In between these lines, nearly-classical  geometry is a useful approximation, so systems can be approximated in the usual way  as quantum fields and particles on a classical background. The model here predicts  new residual effects of quantum geometry that persist on macroscopic length scales; they may exceed standard quantum position uncertainty for transverse positions of massive bodies in the upper half of the figure, above the Planck mass.  }
\end{figure}

It is well known that the general-relativistic paradigm of  dynamical geometry responding to  matter actually becomes inconsistent on scales smaller than the Planck length, $l_P\equiv \sqrt{\hbar G/c^3}$, where $c$ denotes the speed of light, $\hbar$ denotes Planck's constant, and $G$ denotes Newton's gravitational constant (see  Figure \ref{domains}). Beyond that scale,   classical geometry is not a useful approximation\cite{Banks:2011av}.

Many theories address possible new forms of physics at the Planck scale\cite{Hossenfelder:2012jw}.  
Quantum field theory predicts that Planck scale effects are highly suppressed at the much longer length scales accessible with direct experiments,
  but on the other hand,  it {\it assumes} a classical geometry and classical spatial structure for its quantized modes, and this assumption may be incorrect, without conflicting with current experiments\cite{cohen1999,Hogan:2013tza}.
 General states and formal observables in field theory are nonlocal\cite{Giddings:2005id},  so in principle, new forms of entanglement with quantum geometry are possible even in large systems.

 To show that quantum response of geometry to matter is not confined to small scales only, it is useful to
consider the  wave function of a  quantum particle state  that extends over a macroscopic region, for example, the wave function of a single photon that arrives from a distant star, and is captured by a telescope  before it is focused to create a sharp image.
Suppose a photon is radiated from an atom in a typical state, with an indeterminate direction.
In general, the photon state  extends many light-years without having a determinate direction.  
 Now suppose that a measurement, say an absorption by an atom in a detector,  ``collapses'' the location of the photon to a definite, localized position state. The spatial state changes acausally and non locally from a scale of light years to the scale of an absorbing atom. This occurrence is not only instantaneously acausal everywhere at the moment of detection,  it is actually {\it global and retroactive}: the entire historical trajectory previous to the ``collapse event'' is suddenly much more localized than it was before the measurement. In the case of a photon landing on a detector, the entire previous path is localized to the size of the telescopic aperture, as opposed to a region extending many light years.    

Now consider the geometry that accompanies this photon. General relativity couples to classical matter, so there is no standard theory for geometry in this situation. For the geometry to match the gravity of the photon, the geometry, including the accompanying trajectories of all the other matter everywhere determined by it, needs to match the state where the photon's mass is.  Instantaneous, retroactive rearrangement of the geometrical quantum state, as well as the position of all the other matter throughout a macroscopic 4-volume of space and time,  needs to follow the quantum photon wave function.  Quantum  geometry must display  the same  ``spooky'' nonlocal quantum entanglement as  that of the matter that couples to it.  
  
For practical purposes such nonclassical features of geometry can usually be disregarded because the indeterminate gravity of quantum particles is small.     But it is clear that the states of an evolving quantum geometry are nonlocal and macroscopically indeterminate in both space and time, even though any new quantum-geometrical  physics may originate at the Planck scale. Apparently,  the  separation of scales in field theory does not apply to geometry. The model here  captures a particular form of macroscopic indeterminacy and  its distinctive departures from classical trajectories of massive bodies.%, particularly when  quantum measurement  are considered.  

\subsubsection{Macroscopic Quantum Geometry}

Suppose then that
classical spatial relationships,  including  notions of spatial dimension, position, direction, and locality,   emerge as  average approximate behaviors in the macroscopic limit of a  quantum system.  
 The  states of this system can include new quantum-geometrical degrees of freedom and forms of quantum position uncertainty and entanglement that may produce new physical effects  on macroscopic scales,  that are not included in standard  theory.

  A simple model of a macroscopic quantum geometry with these properties is proposed here.
  As a simple approximation, we  dispense with both the conventional gravity and quantum properties of  matter fields altogether, and use an approximation that describes only the information in the  quantum geometry of flat space.  The model is not a fundamental theory, indeed  {\it it  ignores  all standard quantum degrees of freedom}, and cannot describe elementary particles and fields. Instead, {\it it is a model quantum system  that includes only  new quantum  geometrical degrees of freedom.}  Position is represented not as classical coordinates, but as a set of operators  with a  noncommutative algebra.  Basic quantum mechanical principles then
  lead to an model system that predicts new physical behavior: a qualitatively new form of coherent quantum indeterminacy of angular position of massive bodies on macroscopic scales.  
  
  This hybrid of quantum physics and geometry applies  in a regime complementary to that of field theory in a nearly-flat space-time. The mass must be larger than the Planck mass so that standard quantum uncertainty is subdominant, while the separation is large enough that space-time curvature effects  are negligible (see Fig. \ref{domains}).
    
      \subsubsection{Relation to ``Noncommutative Geometry''}

The new geometrical degrees of freedom  are introduced here using a
 noncommutative quantum  algebra of  position operators.  
 Noncommutative geometries have been previously applied to fields and other systems, by deforming the classical background geometry or phase space to a quantum one\cite{connesbook,Douglas:2001ba,bastos}.  The application here differs from those approaches. 
 
The most significant difference is that the noncommutative geometry here refers to an entirely different physical system. The operators here simply refer to positions of otherwise classical massive bodies.  The  model excludes any connection with conventional quantum degrees of freedom. Indeed, the treatment here does not even introduce any dynamics.  No attempt is made to blend the geometry with  fields or other continuous functions on a deformed space.  

In addition, to preserve  covariance in the classical limit,  as explained below, we use  position and 4-velocity operators to contract the 4D antisymmetric tensor to the two indices required for a commutator coefficient.  The 3D commutator  here,   based on contracting the antisymmetric tensor $\epsilon_{ijk}$ with the position of a body,  differs from the standard approach  of combining a two index commutator  $\theta_{ij}$  with an auxiliary field. 
The  self-contained algebra here has fewer degrees of freedom, and increases the geometrical uncertainty.
 
These differences lead to significantly different physical effects. Quantum  positions in this framework do not respect  locality or separation of scales.  They  approximate  classical positions in the macroscopic limit, and approximate locality emerges in a consistent way from the quantum system.  However,
quantum  departures from classical directionality on macroscopic scales, albeit very small, are  much larger in this system than in previous approaches. Because classical geometrical coherence is not enforced by assumption, the macroscopic quantum departures from classical geometry are also more macroscopically coherent.

\subsubsection{Relation to Thermodynamic Gravity}

Internal evidence from gravitational theory has accumulated that  elements of geometry, such as manifolds, events, paths and metrics, may not be fundamental physical entities, and that
 classical properties of  motion--- including  inertia, the principle of equivalence, and gravity--- may be  emergent statistical behaviors.

Early suspicion of a statistical or thermodynamic origin of geometry came from the laws of black hole thermodynamics, derived from the  behavior of classical black hole solutions\cite{Bardeen:1973gs,Bekenstein:1973ur};  for example, the entropy of a black hole was identified with one quarter of  the area of its event horizon in  Planck units. It became more concrete  with the theoretical discovery of Hawking quantum radiation from black holes.\cite{Hawking:1974sw}

Jacobson\cite{Jacobson:1995ab} explained these results and generalized them  by showing that the equations of general relativity themselves can be derived entirely from thermodynamic arguments. He argued that the gravitational lensing  caused by matter must distort the causal structure of space-time in a particular way--- that is, such that the Einstein equations hold---  in order that any accelerated observer see the correct relation $\delta Q= T dS$ between energy flux $Q$, Unruh temperature $T$, and entropy $S$ just inside any Rindler causal horizon.

%``The Einstein equation is derived from the proportionality of entropy and the horizon area together with the fundamental relation $\delta Q= T dS$.  The key idea is to demand that this relation hold for all the local Rindler causal horizons through each spacetime point, with $\delta Q$ and $T$ interpreted as the energy flux and Unruh temperature seen by an accelerated observer just inside the horizon. This requires that gravitational lensing by matter energy distorts the causal structure of spacetime so that the Einstein equation holds.''

%
Thus, general relativity emerges statistically from   states of some holographic quantum system. The  fundamental geometrical  objects in Jacobson's reasoning are not events, but 2D null surfaces.  The key physical effect of curvature  is gravitational lensing, which can also be interpreted (via Fermat's theorem) as a distortion of null wavefronts.

More recently, Verlinde\cite{Verlinde:2010hp}  derived gravity from a statistical theory, but in a different way.  Entropy is again holographic,  and information about position is encoded on two dimensional spacelike surfaces. Emergent behavior of bodies in space, including position, inertia, and  Newtonian gravity,  is derived  from  positional degrees of freedom. The positional relationship between bodies is equivalent to coarse-grained  information on surfaces between them. %In the non relativistic case, they are simply spacelike surfaces. 

In both cases, the derivations  connect one classical theory with another: thermodynamics with gravity. They show that geometry derives from  statistical behavior of some underlying quantum system.
The fundamental degrees of freedom of the geometrical states are not characterized in detail, but the Jacobson and Verlinde treatments highlight some important properties of the quantum system.  The states  appear holographic in the emergent system, and fundamentally respect causal structure. Somehow, the encoded geometrical information   respects  a nearly-classical set of null relationships, while limiting the amount of information overall.

 The states of the quantum geometrical system described below are also nonlocal and holographic, and could serve as a model basis of such a statistical or emergent theory of emergent gravity.  If  the macroscopic states of the underlying quantum system resemble those proposed here,  these classical properties of the system are accompanied by specific new quantum effects.  They create a new quantum-geometrical ambiguity in directional position on 2D spacelike surfaces, while preserving radial separation and causal structure. 
Thermodynamics provide an emergent theory for the motion of  mean position; the quantum geometry here provides a quantum model of a  new  variance from that classical behavior.

 The model here does not extend to curved geometry; in thermodynamic language, it treats only the zero temperature case.
 Although the geometry is a quantum system, there is no  ``quantum gravity'', since curvature degrees of freedom are not quantized.
 However, if we normalize the model to agree with the Planckian  information content suggested by low-energy gravity, the zero-temperature quantum fluctuations can then be predicted with no free parameters. 

\subsubsection{Planck Bandwidth Broadcast}

Previous effective theories with similar macroscopic behavior have been based on the idea that  information about relative position of bodies propagates as a broadcast wave function with a finite Planck  bandwidth\cite{Hogan:2007pk,Hogan:2008zw,Hogan:2010zs}.
The  quantum uncertainty of direction is similar to that in the present model, because in both cases it originates from the limited information in the system.  The uncertainty corresponds to the non-uniqueness  of classical geodesics that describe  a system of bandwidth-limited waves.

The  model proposed here results in slightly different, as well as more specific predictions, because of more explicit and precise constraints from basic principles of covariance and quantum mechanics in the macroscopic limit.  Fewer assumptions are needed to connect the directional uncertainty with the number of degrees of freedom, and thence to thermodynamic gravity; this estimate leads to a somewhat different numerical coefficient for the commutator.  Although the model is connected to new Planckian physics, it need not apply in this form at the Planck scale; rather, it is a  hypothesis, in the form of a quantum mechanical system, about how  attributes of position might depart  slightly from  classical behavior in the macroscopic limit, without reference to any particular microscopic theory.

\subsubsection{Classical Massive Bodies}
The physical quantity identified with the position operator here is the mean position of an assembly of matter, which in keeping with classical use is called a ``body''.  This concept does not have a clear microscopic meaning, but has a clear macroscopic physical interpretation  according to the usual meaning of position in the classical limit. The model  is based on this emergent behavior, rather than a bottoms-up derivation from a fundamental theory.  No attempt is made here to relate quantum geometry to a more fundamental theory, such as string theory, apart from the normalization based on a holographic origin of gravity. 
As shown below, the approach  is self-consistent because the spatial locality used to define the assembly of particles emerges with increasing precision on macroscopic scales; indeed, it can be used as a framework to describe how approximate spatial locality emerges as a property of states of massive systems.

Unlike familiar quantum mechanical systems,  the commutator considered here does not depend on the mass of the body, only on its position. As shown below, the new geometrical uncertainty dominates over standard quantum-mechanical uncertainty for assemblies of matter above a few Planck masses.
The model here thus  serves as a self-consistent approximation for a quantum theory of position in systems that are both much larger than a Planck length in size, and much larger than a Planck mass in total energy.  It cannot serve as a theory of elementary particles, but may be a better approximation than quantum field theory for directional degrees of freedom in large systems (see Figure \ref{domains}).

 \section{Model}

  \subsection{Covariant Commutation Relations}
 
 Suppose that the position of  a  body in each direction $\mu$ is a quantum observable, represented by a self-adjoint operator ${\hat x}_\mu$.  The commutators of these operators represent the quantum deviations of a  body from a classical trajectory. The body  is assumed to be massive enough that we  ignore the conventional position commutators--- the usual quantum effects associated with its motion.
 
Now consider the following candidate quantum commutator relating position operators  in different directions:
\begin{equation}\label{covariant}
[{\hat x}_\mu,{\hat x}_\nu]=  {\hat x}^\kappa  {\hat U}^\lambda \epsilon_{\mu\nu\kappa\lambda}  i\ell_P,
\end{equation}
where indices $\mu,\nu,\kappa,\lambda$ run from 0 to 3 with the usual summation convention,  
$ {\hat U}$ represents an operator with  the same form (in the classical limit)  as the dimensionless  4-velocity of the body, and $\epsilon_{\mu\nu\lambda\kappa}$  is the antisymmetric 4-tensor. 
In the limit $\ell_P\rightarrow 0$, the commutator vanishes, and positions in different spatial directions behave independently and classically.
If the scale  $\ell_P$ is not zero, there is some intrinsic geometrical uncertainty.%,  $\ell_P\approx ct_P$. % The information density of geometrical states then agrees with  holographic entropy bounds for gravitating systems, as explained below. 

The explicit dependence  adopted in Eq.(\ref{covariant}) on  position and velocity is  driven by the requirement of manifest covariance.
The quantum commutator of two vectors requires two antisymmetric indices that must be matched by indices  on the right side.
 Thus we require a nonvanishing antisymmetric tensor, which in four dimensions has  four indices, $\epsilon_{\mu\nu\lambda\kappa}$.  Two of its antisymmetric indices match those of the noncommuting positions.  The other two must contract with two different vectors to avoid vanishing. The  geometrically defined options adopted in Eq. (\ref{covariant}) are the 4-velocity and position of the body being measured.

Equation (\ref{covariant}) is thereby constructed to be manifestly covariant:  the two sides transform in the same way under the homogeneous Lorentz group, as a direct product of vectors.  The algebra of the quantum  position operators  respects the transformation properties of corresponding coordinates in  an emergent classical Minkowski space-time, in a limit where the operators are interpreted as the usual space-time coordinates. 
The model   thus defines no preferred direction  in space.

On the other hand,  Eq. (\ref{covariant}) is not translation invariant. The commutator does depend on the position and 4-velocity of the body being measured, or equivalently, on the origin and rest frame of the coordinate system.  We interpret this to mean that {\it  the  commutator describes a quantum relationship between world lines that depends on  their relative positions and velocities,  but not on any other properties of the bodies being compared}.    The Hilbert space of the  model  quantum geometry  depends  on a causal structure defined by  a choice of timelike trajectory\cite{Banks:2011av}, and the quantum-geometrical position state is defined in  relation to a particular world line,  the origin of the coordinates.  Unlike a classical geometry, which is  defined independently of any observer, the state of a quantum geometry and indeed the definition of position are shaped by this choice, so they cannot obey translation invariance.  As discussed below, if  $\ell_P$ is of the order of the Planck length, this subtle quantum effect would have escaped detection.

A potentially serious problem with Eq. (\ref{covariant}) is that quantum operators  $ {\hat U}^\lambda$ are not well defined.
We can write an expression
$ {\hat U}^\lambda\equiv  \dot  {\hat x}^\lambda(\dot {\hat x}_\alpha \dot {\hat x}^\alpha )^{-1/2}$  with  the same form (in the classical limit)  as the dimensionless  4-velocity of the body, normalized to preserve reparametrization invariance, where $\dot x\equiv \partial x/\partial \tau$ and  $\tau$  denotes proper time. However,  the operator equation should use only the $\hat x_0$ coordinate, not  a classical proper time.  Although Eq. (\ref{covariant}) has a manifestly covariant classical interpretation, we have not proven that it is a consistent operator equation.

We now proceed to show that for systems with near-vanishing 4-velocity, there is a 3-dimensional version of the commutator that is a demonstrably consistent quantum algebra.  Although  it does not provide a fundamental theory, it is sufficient for estimating the spatial uncertainty of transverse position.

\subsection{Small Velocity Expansion}

To show that there is a consistent theory at low velocity,  expand explicitly the components of Eq. (\ref{covariant}).  The  components of classical 4-velocity are $U=\gamma (1, u_i)$, where $\gamma\equiv (1-u_iu_i)^{-1/2}$ and $  u_i\equiv  \partial  x_i/\partial x_0\equiv \partial_t x_i$. Then Eq. (\ref{covariant}) breaks into two parts,
\begin{equation}\label{spacepart}
[\hat x_i, \hat x_j] = i \gamma ( \hat x^k \epsilon_{ijk} + \hat t \partial_t \hat x^k\epsilon_{ijk})
\end{equation}
and
\begin{equation}\label{timepart}
[\hat x_i, \hat t]= i \gamma \hat x^j\partial_t\hat x^k\epsilon_{ijk},
\end{equation}
 where   indices $i,j,k$ now run from 1 to 3, and   $\hat t\equiv \hat x_0$. Note that we have again  adopted a classical time  derivative, without  explaining its relation to the internally referenced operator quantity, $\hat t$. The same approximation was used to describe the classical limit of $U^{\lambda}$, but the operator elements of the manifestly covariant commutator (Eq. \ref{covariant}) have not been been rigorously defined.  
 These approximations do not affect the results below, which are based on a time-invariant, 3D commutator.

The appearance of a ``time operator'' $\hat t$ means that in spite of manifest covariance,  timelike and null surfaces are defined by the operators are frame-dependent.  In Eq. (\ref{spacepart}), an explicit dependence on time coordinate appears, proportional to the   velocity. In Eq. (\ref{timepart}), a nonzero transverse component of $\partial_t\hat x^k$ results in a nonzero  commutator of time with  the radial component of  spatial position, and hence a quantum ambiguity of causal structure. The physical interpretation of this is not clear.

 However, both of these terms  include the 3-velocity, so at low velocity the ambiguity can be neglected. 
For velocities much less than unity, as assumed here for  massive bodies,  the magnitudes of  the problematic terms are  subdominant compared with the  first term in Eq. (\ref{spacepart}), which depends only on  position.  The physical interpretation of the velocity-dependent terms, and the consistency of a fully 4D operator theory, are potentially interesting, but will not be considered further in this paper.

\subsection{Rest Frame Limit}

The physical interpretation of Eq.(\ref{covariant}) is most straightforward when converted into a quantum geometry of spatial position in the classically defined rest frame of a body,  $\partial_t x_i\rightarrow 0$. 
 In that frame, the classical 4-velocity  is  $ U^\lambda= (1, 0, 0, 0)$ so the  non-vanishing terms of Eq. (\ref{covariant}) are those multiplied by $\epsilon_{\mu\nu\kappa\lambda}$ with $\lambda=0$.  The remaining terms  describe a noncommutative geometry in three dimensions:
\begin{equation}\label{3Dcommute}
[{\hat x}_i,{\hat x}_j]=  {\hat x}_k \epsilon_{ijk} i\ell_P.
\end{equation}
  Eq. (\ref{3Dcommute})   describes a quantum-geometrical relationship between positions of two trajectories (or massive bodies) that have   proper 3-separation
$ {\hat x}_k$,  and whose world lines have the same  4-velocity.

The algebra in Eq. (\ref{3Dcommute})   defines a consistent quantum theory; indeed, it is the same as the standard quantum description of angular momentum, albeit in an unconventional physical application. Here, position in units of a fundamental length $\ell_P$ takes the place usually assigned to angular momentum in units  of $\hbar$. The  algebra itself is of course well known\cite{LL}, and leads to precise results on the properties of the quantum-geometrical system.

The operators are well known to obey the Jacobi identities
\begin{equation}
[{\hat x}_i, [{\hat x}_j,{\hat x}_k]] + [{\hat x}_k, [{\hat x}_i,{\hat x}_j]] + [{\hat x}_j, [{\hat x}_k,{\hat x}_i]] =0,
\end{equation}
%(It is also consistent with another new  quantum-geometrical relation,
%\begin{equation}
%[{\hat x}_i,\dot {\hat x}_j]= \dot {\hat x}_k \epsilon_{ijk} i\ell_P,
%\end{equation}
%although this property does not  enter directly into important observable consequences of Eq. $\ref{3Dcommute}$.)
 which  shows that in the non-relativistic  limit (Eq. \ref{3Dcommute}), the effective quantum theory is self-consistent. Note that this conclusion does not require knowing the quantum-geometrical commutators for dynamical operators $\dot {\hat x}_\mu$, or knowing the Hamiltonian for the system. Usually the quantum conditions (Eq. \ref{3Dcommute}) are derived from a correspondence principle  with classical Poisson brackets; here, they are motivated just from their  symmetries and holographic information content in the classical limit, since there is no classical system corresponding to the quantum geometrical degrees of freedom.  
 
 As estimated below, in laboratory applications on a scale of $\approx 10^{36} \ell_P$,   the quantum geometrical transverse positions fluctuate  extremely slowly,  equivalent to about $ 10^{-18} c$, so the nonrelativistic approximation is excellent. More generally, for a body at rest the terms in Eq. (\ref{covariant}) multiplied by $1,2,3$ components of  $ {\hat U}^\lambda$  are smaller in magnitude than the $0$ component by a  factor of order $| x|^{-1/2}$, which is tiny for a macroscopic system, 
 so they can be consistently neglected.  The discussion below is confined to   this limit, and thereby avoids confronting the correspondence of $ {\hat U}^\lambda$ with its classical counterpart, or the subtle  relationship  of  the time operator, $\hat x_0$, with the  classical time coordinate.
 
 \subsection{Radial Separation Eigenstates }

Consider an operator
\begin{equation}
|{\hat x}|^2\equiv {\hat x}_i{\hat x}_i,
\end{equation}
corresponding to the squared modulus of separation, analogous to the square of total angular momentum.
In the same way that total angular momentum commutes with all of its components,  the radial separation between trajectories behaves classically:
\begin{equation}
[|{\hat x}|^2, {\hat x}_i] = 0,
\end{equation}
so that emergent causal structure, with light cones defined by $|{\hat x}|^2$, has no quantum uncertainty.  For any single direction--- for example, a direction of a plane wave mode in field theory--- the system behaves classically. It is only in comparison of positions in transverse directions that new quantum-geometric effects appear.

We adapt conventional notation used for angular momentum. Let $l$ denote positive integers corresponding to the quantum numbers of radial separation, analogous to total angular momentum. The separation operator  takes discrete eigenvalues: 
\begin{equation}\label{radialeigenvalues}
|{\hat x}|^2|l\rangle =l(l+1) \ell_P^2|l\rangle.
\end{equation}
We denote the  discrete  eigenvalues corresponding to classical separation  by
\begin{equation}\label{separation}
L\equiv \sqrt{l (l+1)}\ell_P.
\end{equation}

\subsection{Directional Eigenstates and  Uncertainty}

 A body can be in a  state of definite radial separation, and also definite position in any single direction.
However, a body cannot be in a definite position state in more than one direction, and this leads to a new kind of uncertainty that can also be understood from the angular momentum algebra. 

If a system is in a state of definite angular momentum in one direction, the components of angular momentum in the transverse directions are indeterminate. Similarly, if a body is in a state of definite separation in one direction, the position components in transverse directions are indeterminate--- in the macroscopic limit, very slightly so.

Now consider projections of the operator ${\hat x}_i$.
Let  $l_i$ denote the eigenvalues of position in direction $i$. 
In a state $|l\rangle$ of separation number $l$, the position operator ${\hat x}_i$ can have eigenvalues  in  units of $\ell_P$,
\begin{equation}\label{positioneigenvalues}
l_i= l, l-1, \dots , -l,
\end{equation}
giving $2l+1$ possible values. 
In an eigenstate with a definite value of position in direction $i$,  
\begin{equation}
{\hat x}_i|l, l_i\rangle= l_i \ell_P |l, l_i\rangle.
\end{equation}

We now calculate the variance for the position wave function in transverse directions.
Define raising and lowering operators for each direction:
\begin{equation}
{\hat x}_{3\pm}\equiv {\hat x}_1\pm i {\hat x}_2,
\end{equation}
with equivalent  expressions for  cyclic permutations of the indices.
These can be used to show\cite{LL} that  for any $i$, %LL pp 85-88
\begin{equation}
|{\hat x}|^2= {\hat x}_{i+}{\hat x}_{i-}+ {\hat x}_i^2+{\hat x}_i= {\hat x}_{i-}{\hat x}_{i+}- {\hat x}_i^2+{\hat x}_i.
\end{equation}
Direct calculation (e.g., ref.\cite{LL}) then leads to  the following product of amplitudes for  measurements of  either of the   transverse components  ${\hat x}_j$, with $j \neq i$:
\begin{equation}\label{transverse}
\langle l_i | {\hat x}_j | l_i-1\rangle  \langle l_i-1 | {\hat x}_j | l_i \rangle = (l+ l_i) (l-l_i+1)\ell_P^2/2,
\end{equation}
again for any $i$.

The left side of equation (\ref{transverse}) can be interpreted as the expected value for the operator
\begin{equation}\label{transvariance}
{\hat x}_j | l_i-1\rangle  \langle l_i-1 | {\hat x}_j 
\end{equation}
for components with $j \neq i$, in a state $|l_i\rangle$ of definite ${\hat x}_i$.
For  $l>>1$ and $l_i\approx l$,   this corresponds to the expected variance in  components of  position transverse  to separation:
\begin{equation}
\langle {\hat x}_j | l_i-1\rangle  \langle l_i-1 | {\hat x}_j\rangle \rightarrow \langle {\hat x}_j^2 \rangle.
\end{equation}
We conclude that   position $  {\hat x}_\perp $ in any direction  transverse to separation is indeterminate, with a   variance given by  the  right hand side  of Eq. (\ref{transverse}) in the limit of $l>>1$:
\begin{equation}\label{perpvariance}
\langle {\hat x}_\perp^2 \rangle   =  L \ell_P.
\end{equation}

Transverse positions  show the same kind of quantum indeterminacy as transverse components of angular momentum in the nearly-classical, large-angular-momentum limit.  The physical analog is  measurement of a  component transverse to the angular momentum vector; such a measurement reveals a superposition of transverse direction states, with a range of  eigenvalues.
In our application, position can similarly have a  definite value in the radial direction---  corresponding to its classical value--- but is  indeterminate in the transverse directions. The new quantum-geometrical uncertainty  increases with  separation, and at macroscopic separations, it is much larger than $\ell_P$.

As usual in quantum mechanics, the state can be represented by a wave function, and the commutator leads to an uncertainty relation between the variances of the wave functions of the conjugate variables.
It can be derived from Eq.(\ref{3Dcommute})  in the usual way, although the conjugate variables are now positions in different directions, instead of angular momentum components in different directions. 
In the rest frame,  the uncertainty relations for  a body at  positions ${\hat x}_k$  are 
\begin{equation}\label{3Duncertainty}
\Delta x_i\Delta x_j \ge   | \bar x_k \epsilon_{ijk} | \ell_P /2,
\end{equation}
where $\Delta x_i= \langle |x_i- \bar x_i |^2\rangle^{1/2}$ represents the spread of the wave function in each direction about the expected (classical) position $\bar x_i$. Note that this result is  less constraining than the more powerful result  (Eq. \ref{perpvariance}) derived from the full operator algebra, which shows that uncertainty in this system cannot be squeezed out of one transverse direction and into the other.

\subsection{Total number of eigenstates}
 
 The geometrical position states  have a discrete spectrum.
The   number of position eigenstates  within a 3-sphere of radius $R$ can be counted in the same way as discrete angular momentum eigenstates, as follows.

  Recall that radial position has discrete quantum numbers $l$.
 Using Eq. (\ref{separation}) with $R=L$, the number $l_R$ of radial position eigenstates for a radius $R$  is given by setting $l_R (l_R+1)= (R/\ell_P)^2$, that is, they are discrete with approximately Planckian separation.  For each of these, from Eq. (\ref{positioneigenvalues}), there  are $2l +1$  eigenstates of direction.   The total number of   quantum position eigenstates in a  3-sphere  is then 
\begin{equation}\label{3Dsphere}
{\cal N}_{Q3S}(R)= \sum_{l=1}^{l_R} (2l+1) = l_R(l_R+2)=  (R/\ell_P)^2,
\end{equation}
where the last equality applies in the large $l$ limit.
Thus,  
 the number of  quantum-geometrical position eigenstates in a volume   scales holographically, as the surface area in Planck units.%,  instead of  3-volume, as it does for  the eigenstates of a quantized field with a frequency cutoff.    

\subsection{Normalization from  Gravity}

 Although gravity does not enter explicitly into any of the arguments here,
 it is natural to fix the  physical value of  $\ell_P$ by equating the number of geometrical  eigenstates with the number of states deduced from thermodynamic arguments  applied to gravitational systems\cite{Jacobson:1995ab,Verlinde:2010hp}.  
An exact calibration is possible, because the number of states in a 3-sphere  can be counted exactly for both  geometry and gravity. 
 
 The number of position states for a massive body enclosed in a 3-sphere of radius $R$ that statistically reproduces nonrelativistic Newtonian gravity (cf. \cite{Verlinde:2010hp}, Eq. 3.10) is given by:
  \begin{equation}\label{verlindegravity}
{\cal N}_{G3S}(R)= 4 \pi (R/ct_P)^2, 
\end{equation}
with the usual definition of Planck time, $ct_P\equiv \sqrt{\hbar G/c^3}= 1.616\times 10^{-35}$m. Note that this is four times larger than the entropy of a black hole event horizon of the same radius. 
The quantum-geometrical and gravitational estimates agree, ${\cal N}_{Q3S}={\cal N}_{G3S}$, if  the  numerical value of the effective commutator coefficient $\ell_P$  is
 \begin{equation}\label{value}
\ell_P= ct_P/\sqrt{4\pi}.
\end{equation}
If Eq. (\ref{value}) holds, bodies in the space-time that emerges from the quantum
geometry  move as described by Newtonian  gravity.

 As discussed below, an interesting and attainable experimental goal is to detect or  rule out a  commutator with this value.  From Eq. (\ref{perpvariance}), the theory predicts quantum-geometrical uncertainty in physical units,
  \begin{equation}\label{exact}
\langle {\hat x}_\perp^2 \rangle=   L ct_P/\sqrt{4\pi}= (2.135 \times 10^{-18} {\rm m})^2 (L/{\rm 1 m}) ,
\end{equation}
with no free parameters. Experimental consequences are discussed below.

The numerical value of the coefficient in Eq. (\ref{exact})  is slightly lower (by a factor of $\sqrt{\pi}/4$) than the value previously estimated by a different method, using a  wave theory normalized to black hole entropy (Eq. 29 of ref. \cite{Hogan:2010zs}.) The value computed here is better controlled, because  the quantum and gravitational numbers  both refer to  the same physical system, a spherical 3-volume at rest with no curvature. This value  is the main new quantitative result of this paper.

\section{Emergence of Locality}    
The  effect of the quantum geometry can be characterized as an uncertainty in angular position or direction. Consider the quantum-geometrical variance in angular position of a body at separation $L$, from Eq. (\ref{perpvariance}):
\begin{equation}\label{direction}
\langle \Delta \theta^2\rangle \equiv  \langle {\hat x}_\perp^2 \rangle/ L^2  =  \ell_P / L.
\end{equation}
Unlike the position uncertainty, the directional uncertainty decreases with separation.
Classical  geometry and locality thus emerge  as  excellent approximations to describe relative positions of trajectories with separations much larger than the Planck length. 

For any experimentally accessible scale, the  deviation from classicality is fractionally negligible.
However, as separations approach  the Planck scale,   directions become  indeterminate.   The classical approximation to geometry then breaks down, consistent with the idea of a space-time emerging from a Planckian quantum system. 
That system does not  resemble  a traditional quantum-gravitational ``space-time foam'' with quantum equivalent of strong Planck scale gravity and virtual black holes, inspired by extrapolation of quantum field theory. Instead, positions, curvature and gravity  only  emerge as  statistical behaviors on large scales.

\subsection{Angular Fluctuations and Coherence}
The  uncertainty  describes properties of a wave function in a stationary state that respects the symmetries of a body at rest.  As usual,  the   wave function itself  is  not observed  in any physical measurement, since it describes a superposition of histories and position states.

In the proposed model, the static wave function has the following  physical interpretation:  the spread of the wave function  manifests as random results, or noise,  in a time series of nonlocal position measurements in more than one direction.  
The effect of the uncertainty on measurements of position is constrained by requiring the system  to approach standard physics in the macroscopic limit.

\begin{figure}%[htbp]
 \epsfysize=2.5in 
\epsfbox{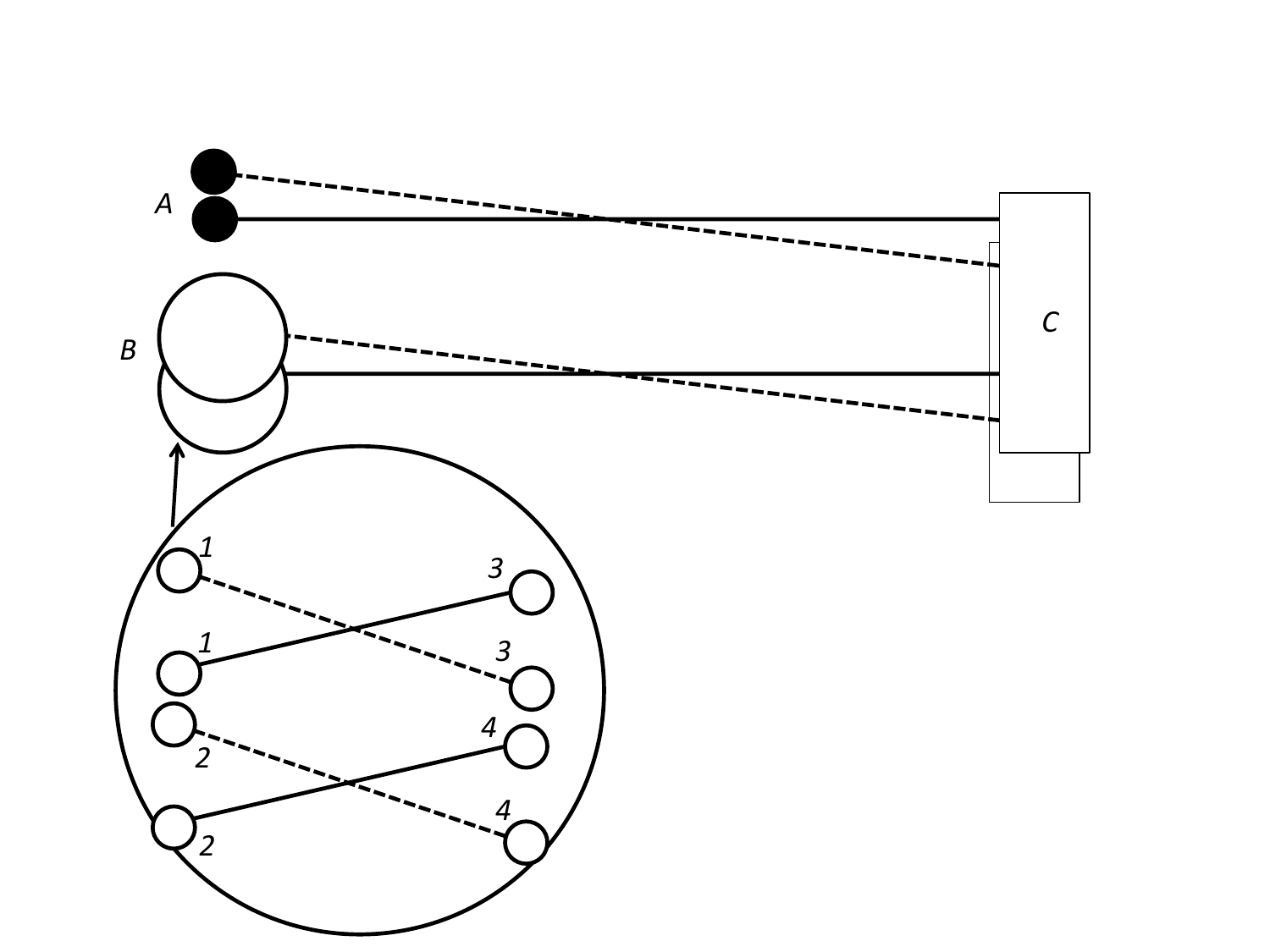} 
\caption{ \label{recursive}
Illustration of how locality can emerge via recursive hierarchical  entanglement.  
Causally connected events share the same geometrical state: ÒcollapseÓ of a transverse position state wavefunction spreads on a light cone from each event where a null surface that defines a measurement in a particular direction intersects a surface of constant time, in the rest frame of the body used to measure the system. This idea can be used to recursively extrapolate the behavior of a single-body state to a system of multiple bodies. 
A massive body is defined as  a collection of smaller ones in a localized region, and its position is defined relative to distant regions.
  Consider  a pair of nearby bodies, $A$ and $B$, and their direction relative to  a distant body $C$.  The transverse separation is small,  $AB <<AC \approx  BC$.  The $AC$ direction and $BC$ direction share the same geometrical state and fluctuate coherently on timescale $\approx AC\approx BC$. Two possible orientations are shown as solid and dashed lines. Similarly, consider components of $B$, labeled 1, 2, 3, and 4. Their positions fluctuate coherently relative to $C$ on the timescale $BC$. They also fluctuate relative to each other on the shorter timescale $13 \approx 24<<BC$. Again, two possible orientations are shown.  The directional fluctuation is larger for the subsystem but the transverse motion is smaller.  A  quantum system with these properties gradually approximates classical locality on large scales. However, its spatial information content grows only holographically with scale.
}
\end{figure}

The state described by the geometrical wave function  cannot  independently describe the trajectories of  individual particles, since that would result in unacceptable, locally measurable fluctuations dependent on an arbitrary choice of origin for the coordinate system. However, the uncertainty is consistent with emergence of locality on large scales,  if the quantum-geometrical   deviations from a classical trajectory   are entangled coherently and non-locally, so that all matter in a small region (where ``small'' means much smaller than  $L$) shares approximately  the same deviation $x_i- \bar x_i$ in an actual position measurement. The scaling of geometrical uncertainty with $L$ then allows a hierarchical recursive emergence of locality on large scales, as illustrated in Figure (\ref{recursive}).

Colloquially, we can say that matter which is close together, moves together. In a Copenhagen interpretation of quantum mechanics, one would say: {\it The geometry ``collapses'' into the same state for   events on the same null surfaces defined by the causal diamond of the observer.} Then, in spite of the observer-dependent branching of the wave function, classical causal structure emerges as a symmetry of the macroscopic system.

This model is consistent  with the classical approximations used to interpret  Eqs. (\ref{covariant}): the quantum operators refer to a collective position and 4-velocity of an arbitrary assembly of  mass-energy in a compact region of space, and  average over time to the mean classical space-time position, as long as the region is much smaller than $L$.

In this model, the positional quantum states of bodies in emergent  space-time  possess a new kind of  nonlocal coherence not describable by states  of standard quantum theory in classical space-time.  The coherent entanglement of geometrical position states creates a new  correlation in the mean positions of otherwise separate bodies--- including an in-common, coherent quantum-geometrical deviation from their classical trajectories. There  is no such thing as a massive body exactly at rest, but a world line (and  rest frame) becomes directionally better defined on large scales.
The decoherence of position states in space and time--- the deviation from noiseless classical behavior--- is  given by the  uncertainty amplitude  in Eq. (\ref{perpvariance}), where $L$ corresponds to the separation of the elements and $L/c$ the timescale of the variation. Put another way, the direction to a distant body, together with its neighbors, fluctuates  slowly about its classical value with variance given in Eq. (\ref{direction}). 

This interpretation of the  uncertainty  opens up a way to build an experiment that probes Planck scale physics.
Positions  fluctuate, with a power spectrum of  angular variations given approximately  by the Planck time--- that is, in an average over duration $\tau$, the mean square variation is
\begin{equation}
\langle\Delta \theta^2\rangle_\tau= \ell_P/c\tau.
\end{equation}
In an experiment of size $L$, the variations accumulate up to durations $\tau=L/c$, ultimately leading to a total variance in position given by the overall uncertainty, Eq. (\ref{exact}).
This prediction can be tested by making very sensitive measurements of  transverse positions of massive bodies. 

 To detect the effect, an apparatus must  compare positions of compact bodies distributed over an extended region  of space in more than one direction, over a duration comparable to its size. Over longer durations, the fluctuations are smaller, as positions in a compact region average to their classical values. 

Fluctuations should appear in a time series of such  comparative measurements.  In the time domain, the  covariant formulation (Eq. \ref{covariant}) contains an operator $\hat x_0$, that approximately corresponds to standard coordinate time. However, this correspondence is not exact.  In the model of emergent locality presented here, a local clock is associated with a particular direction in space. It can be synchronized perfectly in the standard way\cite{wigner,salecker} with other clocks in that direction at large distances, using light to compare clock ticks. However, it displays Planckian noise when compared with clocks at macroscopic distances in other directions. As discussed below, the magnitude of the effect is currently unobservable in tests on a cosmic or microscopic  scale, but might be detected as a new source of noise in interferometer signals.

\subsection{Quantum States and Entanglement}

Nonlocal entanglement of position states is perhaps the strangest  feature of this quantum geometry, but appears to be required in order that the number of position states not exceed the holographic gravitational bound.
Because of the coherence, neighboring   elements of an apparatus  fluctuate together, even if there is no physical connection between them apart from relative  proximity. 
 It is worth sketching how the entanglement works in a somewhat more formal way.

The full quantum  state includes all possible combinations of positions in what eventually appears, to an emergent observer, as  a macroscopic spatial volume of size $L$.  It includes a superposition of all the combinations of discrete choices of position.
The basis states are formed around a particular world line, the origin of classical coordinates. A special role is assigned to the radial quantum number $l$, which is associated with the emergent causal structure that defines that particular world line. The angular quantum number can be represented as a projection associated with any direction in the emergent space. 

Based on the algebra above, let  us write down a position state of a single massive body $A$ at rest  on the $z$ axis. Again, this description will ignore the standard quantum physics attached to $A$, and treat only the quantum-geometrical part.
The natural basis to choose is the radial direction to the body from the origin.
Any  state  within separation $L$ can be written as a superposition,
\begin{equation}\label{singlebody}
|{  A }\rangle= \sum_{l=1}^{l(L)}  \sum_{l_z=-l}^{l_z=l} \alpha_{l, l_z}(A)   | l, l_z\rangle 
\end{equation}
where $\alpha(A)$ is the wave function, the complex  amplitude for a system--- a massive body in a quantum spacetime--- to be in each eigenstate.
Each component is labeled by discrete quantum numbers $l,l_z$ that label  corresponding eigenstates.
  Separation  $L$ comes in discrete eigenvalues (from Eq. \ref{separation}) $L/\ell_P=\sqrt{ l(l+1)}$.  The eigenvalues that correspond to direction are described by projection onto any arbitrary direction; in this case we have chosen the $z$ axis. In a state $|l\rangle$ of separation number $l$, the  operator for  position ${\hat z}$ can have eigenvalues  in  units of $\ell_P$,
\begin{equation}\label{positioneigenvalues}
l_z= l, l-1, \dots , -l,
\end{equation}
giving $2l+1$ possible values.

The separation operator commutes with any of the component projections--- separation acts classically--- but positions in different directions do not  commute with each other. 
The state in Eq. (\ref{singlebody}) is written in terms of radial separation and  $z$ component, and has eigenfunctions with definite values of $z$.  Such a state can also be represented by a superposition  of states in other directions, and has an indeterminacy in  transverse position, given  by Eq. (\ref{exact}).

  \subsubsection{Emergence of Locality via Hierarchical Recursive Entanglement}

 The quantum states of multiple bodies are not independent, but are entangled with the same quantum-geometrical system. 
 A composite system of two bodies at rest can be written as a tensor product of subsystems,
\begin{equation}\label{twobodies}
|{A   B}\rangle=
  |{ A }\rangle \otimes |{  B }\rangle
\end{equation}
The states  of  two  bodies close to each other are entangled, so that projections onto transverse components are coherent.
When positions of $A$ and $B$ are both measured in the $z$ direction, they share the same natural decomposition, and they can both be in eigenstates of $z$.
When they are measured in a transverse  ($x$ or $y$) direction in this state, the transverse position measurement yields a definite value, with probability determined from the squared modulus of their transverse wave functions. They are placed in a new state projected onto the transverse direction.
If world lines $A$ and $B$ are close together compared with $L$,  measurements of $x_A$ and $x_B$ (or $y_A$ and $y_B$) are highly correlated.  Because they share almost the same causal structure,  they share the same projected states (see  Figure \ref{recursive}).

\subsection{Geometrical  and Standard Quantum Uncertainty}

 The standard quantum limit \cite{caves1980,caves1980b}  for the variance in a body's position difference measured at two times separated by a duration $\tau$ is:
\begin{equation}\label{sql}
\Delta x_{SQL}^2\equiv \langle (x(t)-x(t+\tau))^2\rangle \ge 2\hbar\tau/m.
\end{equation}
This can be derived either from non relativistic quantum mechanics, where it includes the reaction force on the mass due a measurement, or from quantum field theory in the cavity modes of an interferometer with a beamsplitter of mass $m$, where it includes the quantum correlation of photon occupation numbers and shot-noise pressure fluctuations in the two arms.

For any $L$, the standard quantum uncertainty (Eq. \ref{sql}) is greater than the quantum-geometrical position variance (Eq. \ref{exact}) at separation $L= c\tau$ if the mass $m$ of a body satisfies $m<4\sqrt{\pi} m_P $, where  $m_P\equiv \hbar/c^2t_P=2.176\times 10^{-5}$g denotes the Planck mass. That is, for masses below the Planck mass, the new quantum geometrical uncertainty is negligible compared to standard quantum uncertainty, for systems of any scale or configuration.  This crossover is shown  in figure (\ref{domains}).

\subsection{Scope and Limitations}

It is useful to summarize what this model does and does not do. 

Essentially, it is a model for quantum geometrical degrees of freedom, that only applies to the position of a massive body at rest, at macroscopic separation.  Although the physics of this system are completely trivial in standard physics, the model shows possible new  quantum effects from the geometry, that are not included in standard approximations. 
%The domain where the new effects  dominate standard effects is shown in figure (\ref{domains}).
 A more detailed model of the relationship with quantum field theory\cite{Hogan:2013tza} describes  how field states could be directionally entangled with geometrical states on large scales.

The model  provides a concrete hypothesis about how locality might  emerge on macroscopic scales from a quantum theory. It also provides a concrete hypothesis about the character of the position states in an emergent space-time.  % Their statistical behavior emerges on large scales to approximately mimic classical space-time, including dynamical concepts such as inertia, force, and gravity; those emergent properties have been previously discussed (e.g. \cite{Verlinde:2010hp}).
 The main new useful result is to provide concrete and definite predictions about the character of the new quantum-geometrical effects. The model provides a new,  exact connection between the number of  discrete geometrical degrees of freedom,  and the width of a transverse wave function of position.  These two quantities are  observable in different ways: the first from gravity, the second from measurements of fluctuations in the transverse positions of massive bodies.

At the same time, the model here is extremely limited in scope and applicability. It includes no Hamiltonian, and  no equations of motion. Classical motion and inertia only arise in the emergent system at a macroscopic level. The quantum-geometrical ``jitter'' itself is so slow--- it has so little energy--- that it is dynamically negligible for any system much larger than the Planck scale.
%%The  model  has little direct connection with dynamical Planck scale quantum theories such as  string theory,  or fields propagating in deformed, noncommutative geometries (e.g., \cite{Douglas:2001ba}).
% Indeed, it refers only to a complementary, macroscopic regime.
%It is possible to posit consistent commutators for new conjugate operators
%${\hat {\dot x}}_j$, 
%for example in 3D:
%\begin{equation}\label{3Dxdot}
%[{\hat x}_i,{\hat {\dot x}}_j]=  {\hat {\dot x}}_k \epsilon_{ijk} i\ell_P;
%\end{equation}
%however, the physical interpretation of these is unclear and there may be no observable effects associated with them.

The model in 3D describes only a time-invariant state. It is equivalent to a theory of angular momentum with no Hamiltonian, and no theory of transitions between states. Nevertheless, it is natural to interpret the width of the wave function as a source of noise in a time series of directional measurements.

\section{Experimental Tests}

 \subsection{Particle Experiments} 
 The purely transverse character  implies zero quantum-geometrical uncertainty--- equivalent to a  classical geometry--- for measurements of position that involve interactions with  particles propagating in only a single direction. This behavior is different from fluctuations or dispersion caused by gravitational waves or quantized-metric fluctuations. It applies even over cosmic distances, so the theory  is unconstrained by studies of light from distant sources that constrain some forms of Planck-scale Lorentz invariance violation.\cite{fermi2009,Laurent:2011he,HESS:2011aa} 

%Quantitatively, the   rate of spatial decoherence is given by differentiating Eq. (\ref{3Dcommute}):
% \begin{equation}\label{coherence}
%\partial_k [x_i,x_j]= \epsilon_{ijk} i\ell_P (\partial x_k/\partial \bar x_k).
%\end{equation}

 The predicted fluctuations\cite{Hogan:2010zs}, while  much larger than the Planck length, are also too small to detect with  laboratory position measurements involving modest numbers of atoms or molecules.  Indeed,  standard quantum uncertainty (Eq. \ref{sql}) dominates geometrical uncertainty even up to bodies of order the Planck mass.  
Quantum-geometrical uncertainty   is   negligible on the  mass scale of Standard Model particles ($<10^{-16} m_P$), which accounts for why classical space-time is such a good approximation for systems involving small numbers of  particles, and why standard theory agrees so well with precision tests in microscopic experiments\cite{Hogan:2013tza}. The space-time even in a single $\approx {\rm TeV}^4$ collision volume encompasses $\approx 10^{32}$ degrees of freedom, and detectors do not have sufficient precision to detect the predicted tiny deviations from the  phase space  of quantized field modes in a classical geometry.  Predicted effects   in particle experiments  were estimated in ref.\cite{cohen1999} for similarly constrained modes, and found to be unobservable.

\subsection{Interferometry}

 Detection of the quantum-geometrical uncertainty  requires a nonlocal, bidirectional experiment, that measures relative  positions of  bodies on at least three widely separated world lines, with time sampling comparable to their separation.
It can best be distinguished from other noise sources by measuring mean positions of large assemblies of particles, that is, massive bodies such as mirrors.

 A Michelson interferometer creates a  coherent state of many photons spatially extended along two macroscopic arms\cite{caves1980}. The    measured signal from the interfered light depends  on  positions of three mirrors, a beamsplitter and two end mirrors, at three different times separated by the arm length $L_a$. The  world lines  of the mirrors are widely separated in  two directions, so quantum-geometrical position uncertainty of the beamsplitter relative to the end mirrors leads to noise, or fluctuations in the signal. Quantum geometrical position fluctuations  with  variance given in Eq. (\ref{exact}), and timescale $\tau \approx L_a/c$,  create noise with calculable statistical properties. 

The statistical properties of the noise are estimated in ref.\cite{Hogan:2010zs}.
 In a standard Michelson interferometer, the power spectral density of position noise in conventional units (variance of  fractional arm length difference fluctuation per frequency interval),  at frequencies low compared with $c/L_a$,  is given approximately by the Planck time:
\begin{equation}
h^2\approx {\langle {\hat x}_\perp^2\rangle\over cL_a}= t_P/\sqrt{4\pi}= (1.23\times 10^{-22} {\rm Hz}^{-1/2})^2.
\end{equation}

More precisely, the  time-domain autocorrelation of the noise in the signal stream of an interferometer with arm length $L$, and hence the frequency spectrum of fluctuations, can be related  to the causal structure of the emergent space-time.  Let $X(t)$ denote the  position difference of a beamsplitter between two directions, as measured by the phase of light at the asymmetric port of an interferometer. The noise autocorrelation is given by
\begin{equation}\label{auto}
\Xi(\tau)\equiv \langle X(t)X(t+\tau)\rangle.
\end{equation}
The zero lag value $\Xi(\tau=0)$ corresponds to the width of the wave function, and is given exactly by the absolute value derived here, Eq. (\ref{exact}).
As noted above, the author regards this value  as better motivated than that given in \cite{Hogan:2010zs}.

\begin{figure}%[htbp]
 \epsfysize=2.5in 
\epsfbox{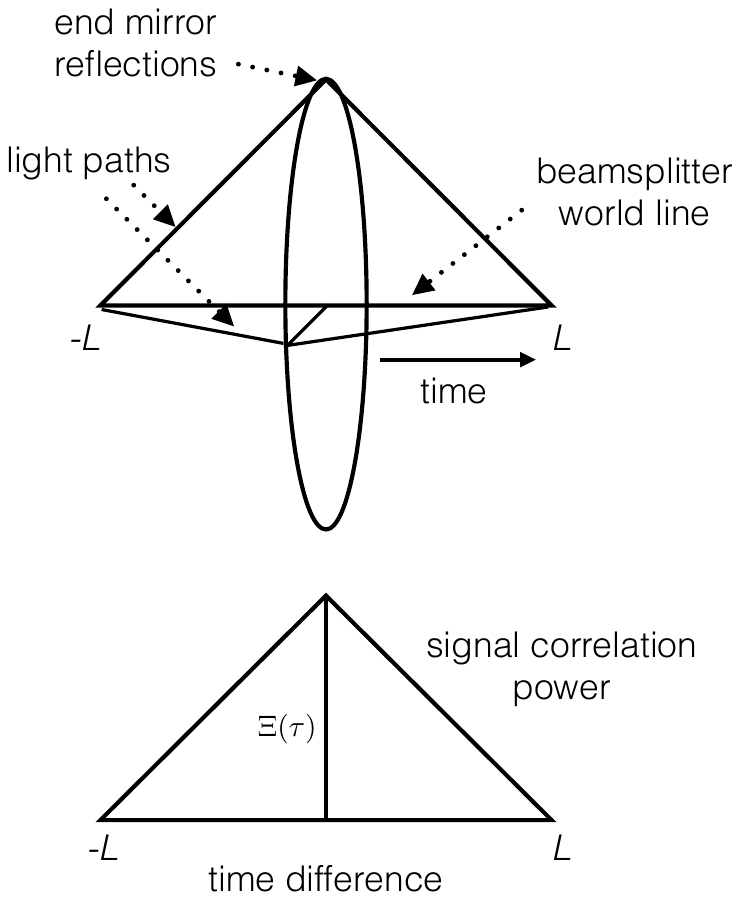} 
\caption{ \label{causal}
Relation of the causal structure of an interferometer to the time domain autocorrelation of the measured phase difference between arms. Above, the causal diamond defined by the two beamsplitter reflection events that enter into the antisymmetric signal for  a simple Michelson interferometer with arm length $L$.  Below, one hypothesis (the most compact) for the corresponding time-domain signal correlation function, Eq. (\ref{auto}).  According to the model  of quantum geometric uncertainty presented here, the value at $\tau=0$ is given by the variance of the transverse position wave function, Eq. (\ref{exact}).}
\end{figure}

The detailed time structure of the signal depends on the causal structure of the space-time inhabited by the interferometer, as shown in Figure (\ref{causal}).
Here, the exact prediction for the time correlation and frequency spectrum depends on the detailed optical layout of the interferometer in space. Predicted spectra for several examples are given in ref.\cite{Kwon:2014yea}.
Care must be taken in using quoted limits from estimates of strain noise, since  the quantum-geometry effect on the signal depends on the interferometer layout in ways that significantly differ from gravitational wave response.  However, it is clear that existing interferometers\cite{GEO2011} already rule out  such fluctuations for commutators significantly  larger than  the Planck scale.

 If two separate interferometers  have  mirrors in nearly the same positions,  separated by much less than  $L_a$, their  geometrical position states are entangled, so  fluctuations in their signals are correlated in a way that depends  on the experimental configuration. 
 The effect of entangled quantum-geometrical position fluctuations can be separated in this way from other sources of  environmental or quantum  noise.

 An experiment based on this specific signature, currently under construction at Fermilab\cite{holometer},  is designed to achieve the required  Planckian sensitivity to  detect the fluctuations, or to convincingly rule them out. The theory here suggests that the outcome of this experiment, either positive or negative, could shed light on the emergent origin of classical locality and gravity from a Planck scale theory.

\acknowledgments
I am grateful to  D. Berman, A. Chou, D. Gross, S. Meyer, J. Schwarz, C. Stoughton, and especially M. Perry, for useful discussions, and for the hospitality of the Aspen Center for Physics, which is supported by  National Science Foundation Grant No. PHY-1066293.  This work was supported by the Department of Energy at Fermilab under Contract No. DE-AC02-07CH11359. 
{}

\end{document}